\documentclass{aa}  
\usepackage{mathrsfs}
\usepackage{amsmath}
\usepackage{graphicx}
\usepackage{txfonts}
\usepackage[multiple,symbol*]{footmisc}
\usepackage{natbib}
\def\apjl{ApJ}                
\def\apjs{ApJS}               
\def\ao{Appl.~Opt.}           
             
\def\aap{A\&A}

\def\acena{$\boldsymbol{\alpha}$~{\it Cen~A}}
\def\acenb{$\boldsymbol{\alpha}$~{\it Cen~B}}
\def\proxcen{{\it Proxima Cen}}

\def\muhz{$\mu$Hz}          
\def\ms{m~s$^{-1}$}         
\def\cms{cm~s$^{-1}$}       
\def\m2s2{m$^{2}$~s$^{-2}$} 
\def\msol{M${_\odot}$}             

\begin{document}
   \title{Asteroseismology of {\acena} \thanks{Based 
   on observations collected with the HARPS spectrograph at the 3.6-m telescope 
	   (La Silla Observatory, ESO, Chile: program 075D-0800A)}}

   \subtitle{Evidence of rotational splitting}

   \author{ M.~Bazot \inst{1,2}, F.~Bouchy\inst{3,4}, H.~Kjeldsen\inst{1}, S.~Charpinet\inst{2}, M.~Laymand\inst{2} \and S.~Vauclair\inst{2}}
   \offprints{M. Bazot}

   \institute{
           Danish Asteroseismology Centre, Institut for Fysik og Astronomi, Aarhus Universitet, Ny Munkegade, Bygn. 1520, 8000 Aarhus C.\\ 
             \email{bazot@phys.au.dk}              
	\and
   	     Laboratoire d'Astrophysique de Toulouse-Tarbes, Observatoire Midi-Pyr\'en\'ees, CNRS, 
   		UMR 5572, UPS, 14 av. E. Belin, 31400 Toulouse, France  
        \and
	      Observatoire de Haute Provence,
	      04870 St Michel l'Observatoire, France
	\and
	       Institut d'Astrophysique de Paris, 98bis Bd Arago, 75014 Paris, France\\
             }

   \date{Received ; accepted}

  \abstract
   {Asteroseismology provides a unique tool for studying stellar interiors. Recently p modes have been detected on the bright solar-like star {\acena} thanks to high-precision radial-velocity measurements. However a better characterisation of these p modes is clearly needed to constrain theoretical models.}
   {We observed {\acena} during five nights using the HARPS spectrograph in order to improve our knowledge of the seismic properties of this star.}
   {We performed high-precision radial-velocity sequences and computed the acoustic spectrum of {\acena}.}
   {We identify 34 p modes with angular degree $l=0-3$ in the frequency range 1.8-2.9 mHz and amplitude range 13-48 {\cms}, in agreement with previous seismic studies. We find an enhancement of the frequency scatter with the angular degree $l$ that indicates, considering the high inclination axis of {\acena}, rotational splitting and explains the low values of previously suggested mode lifetimes. We also derive new values for the small separations that take the effect of rotational splitting into account .
}
   {Our seismic study of {\acena} leads to a list of now well identified p-mode frequencies and shows the importance  of taking the rotation into account in order to properly characterise the p modes even in quite short campaigns.}
   \keywords{stars: individual: {\acena} -- stars: oscillations -- techniques: radial
   velocities}

   \authorrunning{Bazot et al.}
   \maketitle
%

\section{Introduction}

 The $\alpha$ {\it Cen} system is the closest stellar object to the Sun. It is a triple system with one M dwarf, {\proxcen}, in a large orbit around two stars closer to each other, {\acena} and {\it B}. Those are sun-like stars with spectral types, respectively, G2V and K1V.

 With $V = -0.01$ {\acena} is a very bright star. Because it is part of a visual binary, good constraints on its mass, $M = 1.105 \pm 0.007$ \msol, have been obtained \citep{Pourbaix02}. It is thus an ideal target for asteroseismology of sun-like stars. Several attempts to detect p-modes have been made \citep{Gelly86,Pottasch92,Schou01} until \citet{Bouchy02} identified twenty-eight oscillation frequencies using the CORALIE spectrograph. A five-night multi-site run, coupling measurements from UVES and UCLES, led to the confirmation of those p-modes, as well as to the detection of additional ones \citep{Butler04,Bedding04}.

 Modelling {\acena} has proven challenging. It appears that stars with masses around 1.1 {\msol} stand on the fringe of having a convective core -see e.g. \citet{Bazot04} and references therein- which is very difficult to model properly. Thanks to the various available constraints, oscillation frequencies, parallax \citep{Soderhjelm99} and, hence, luminosity, effective temperature, metallicity \citep{Neuforge97}, and radius \citep{Kervella03}, several models have been computed using these constraints \citep{Thevenin02,Thoul03,Eggenberger04,Miglio05,Montalban06}. However, it is difficult to find models that correctly account for all the observations simultaneously as stated in these publications. Furthermore some problems in the modelling of the interior of {\acena} are still partially unsolved, like the possible presence of a convective core. Therefore, better constraints on the p modes are needed.

 We observed {\acena} during a five night run with the HARPS spectrograph. This paper presents the results of this run. We first give an account of the observations (Sect.~2). In Sect.~3, we describe the techniques used to obtain the acoustic spectrum. We also explain the methods used for frequency extraction and identification. Thirty-four p modes were identified and for some of them we identified multiple frequencies. We discuss, in Sect.~4, the observed scatter in the frequencies with respect to the asymptotic formula \citep{Vandakurov67,Tassoul80}. We present evidence that this scatter shows the signature of rotational splitting, which contradicts a previous interpretation involving pulsating modes with short lifetimes compared to the Sun \citep{Kjeldsen05}. New values are given for the small separations that take this effect into account.

\section{Observations and data reduction}

\begin{figure}
\includegraphics[width=5.5cm,height=8.5cm,angle=-90]{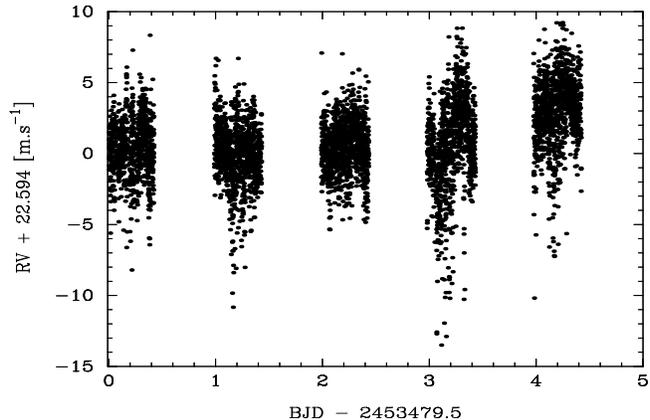}
\caption{
Radial velocity measurements of {\acena}. The dispersion for each individual night is given in Table~\ref{journal}. }
\label{rv}
\end{figure}

\begin{figure}
\includegraphics[width=5.5cm,height=8.5cm,angle=-90]{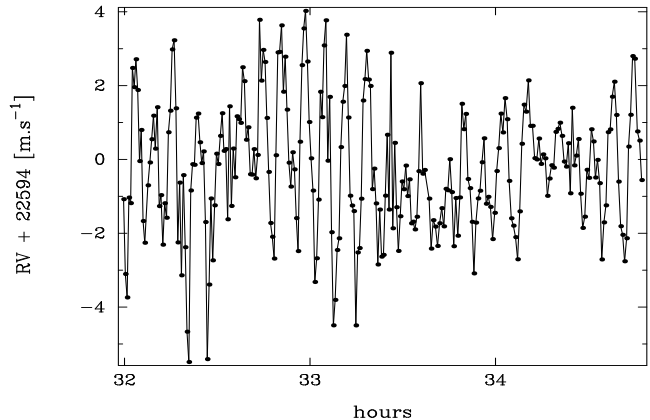}
\caption{Zoom of radial velocity measurements in a three-hour sequence, taken during the third night, showing the presence of p-modes in the time 
series with periods of around 7 minutes. The semi-amplitude of about 3 {\ms} does not 
represent the individual amplitude of p-modes but comes from the interference of several 
modes.}
\label{rvzoom}
\end{figure}

 The target was observed over 5 nights in April 2005, using HARPS at the ESO 3.6 m telescope. We took sequences with typical exposures between 2 s and 10 s, with a dead time of 31 s in between. In total, 4959 spectra were collected with typical signal-to-noise ratios ($S/N$) per pixel in the range 300-450 at 550 nm. The obtained spectra were extracted on-line and in real time using the HARPS pipeline. Wavelength calibration was performed with ThAr spectra, and the radial velocities obtained by weighted cross-correlation with a numerical mask constructed from the Sun spectrum atlas. The journal of the observations is given in Table 1 and the radial velocities are presented in Fig.~\ref{rv}. Following \citet{Bouchy01b} we computed the photon noise uncertainty on the radial velocity and found an average value of 18.5 {\cms} per measurement. The dispersion for each individual night is in the range 1.5-3.3 {\ms}, far from the photon noise value, and is dominated by acoustic modes with periods around 7 minutes, as shown in Fig.~\ref{rvzoom}. Guiding errors, especially during nights 4 and 5, also introduce a non negligible source of noise as discussed in the next section.

\begin{table}
\caption{
Journal of radial velocity measurements. The minimum and maximum $S/N$ of spectra 
obtained each night is given. $\sigma_{{\mathrm RMS}}$ represents the dispersion, expressed in {\ms}.
}
\label{journal}
\begin{center}
\begin{tabular}{ccccc}\hline \hline
Date & Nb spectra & Nb hours & $S/N$ & $\sigma_{{\mathrm RMS}}$ [{\ms}] \\
\hline
2005/04/18 & 835  & 10.05 & 150-550 & 2.04 \\
2005/04/19 & 1040 & 10.59 & 150-500 & 2.10 \\
2005/04/20 & 1038 & 10.63 & 250-500 & 1.80 \\
2005/04/21 & 992  & 10.86 & 100-500 & 3.26 \\
2005/04/22 & 1054 & 10.69 & 200-500 & 2.44 \\ \hline
\end{tabular}
\end{center}
\end{table}

\section{Acoustic spectrum analysis}

\begin{figure}
\includegraphics[width=5.5cm,height=8.5cm,angle=-90]{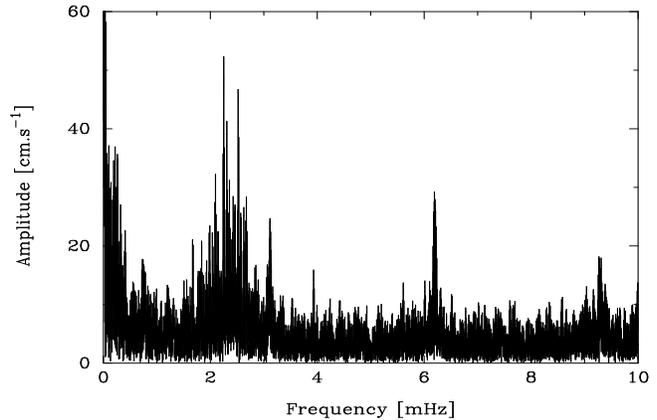}
\caption{Amplitude spectrum of the radial velocity of {\acena}.}
\label{spectrum1}
\end{figure}

To compute the power spectrum of the velocity time series displayed in Fig.~\ref{rv}, 
we used the Lomb-Scargle modified algorithm \citep{Lomb76,Scargle82} for unevenly spaced data. The resulting LS periodogram, shown in Fig.~\ref{spectrum1}, exhibits a series of peaks between 1.6 and 2.9 mHz modulated by a broad envelope, which is the typical signature of solar-like oscillations. This signature also appears in the power spectrum of each individual night. The power spectrum also exhibits three peaks at 3.1, 6.2, and 9.3 mHz.

\subsection{Guiding noise}
These peaks result from a recently identified guiding noise. This is caused by the telescope motor experiencing a hard point during its rotation, which has a period of approximately 5.5 mn. This problem affects the tracking system, causing the stellar light beam to move inside the optical fibre, without significant changes of flux. Such effects should be corrected by the guiding system. Unfortunately this can be done only if the exposure time is longer than the duration of one guiding cycle, approximately 2 s, which was not the case during our run. Furthermore, the guiding camera exposure time, for very bright stars like {\acena}, is as short as 0.01 second. With such a short exposure time and with good seeing conditions, speckles become visible and the guiding system no longer tracks the Gaussian centre of the stellar light spot seen by the fibre, but the brightest speckle. Within these conditions the HARPS guiding system is not well-suited, so it provokes large displacements of the stellar spot inside the fibre. This behaviour of the guiding system is the reason for the radial velocity spikes we can see in Fig.~\ref{rv} during the last two nights. They correspond to the best seeing conditions of the run. The field of view of the fibre is 1 arcsec. In terms of radial velocity, the aperture of the fibre at the entrance of the spectrograph corresponds to a slit of width 3 km.s$^{-1}$. During commissioning tests, the typical scrambling was estimated around 400, which means that a spot moving from the centre to the edge of the fibre would induce a change in radial velocity of about 8 m.s$^{-1}$. The magnitude of this effect decreases with the increasing size of the spot. With a 1 arcsec seeing, a 3-4 m.s$^{-1}$ effect on the radial velocity was measured during commissioning. In nominal cases, the RMS stability of the guiding system is about 0.1 arcsec, and the guiding noise is then estimated to 30-40 {\cms}. This noise is considerably reduced, and eventually vanishes, for long exposure times ($\ge$5 mn), when the correction due to the guiding system cycle (2 s) is completely averaged. This was obviously not at all the case during our run. Especially during the two last nights, we clearly see several spikes with amplitudes from 4 to 8 {\ms} in the RV times series. In the power spectrum the guiding system should introduce a significant contribution to the noise.  

Considering Gaussian noise, we found a 3.7 {\cms} mean noise level in the amplitude spectrum in the frequency range 4-5.5 mHz. Since the time series is based on 4959 measurements, the corresponding velocity accuracy is 1.47 {\ms}. This value is far from the estimated photon noise of 18.5 {\cms}. To verify this value, we separated the contributions to the radial velocity coming from the odd and even spectral orders of the spectrograph. Because there is no dependence of the radial velocity on the parity of the order, subtracting the two curves will suppress all Doppler signals, including the contribution to noise due to the guiding. The remaining signal can be considered as pure photon noise. In the Fourier space we measure a noise of 0.51 {\cms}, indicating a photon noise of 20.3 {\cms} for the whole spectrum, in very good agreement with the first estimate.

We can also compare this measurement to those issued from the HARPS run on $\mu$ Ara \citep{Bouchy05}. The noise level was 1.17  {\ms}, which is significantly lower. But in that case the exposure time was 100s, allowing for the guiding noise to be averaged, at least partially.
Consequently we may conclude that the noise level of 1.47 {\ms} comes in large part from the guiding noise and is seven times larger than the photon noise. This analysis addresses the problems one can encounter when observing bright stars with HARPS. It is necessary to use exposure times long enough to average the guiding contribution to the noise. In case this is not possible, as for {\acena} (because saturation would be reached with longer exposure times), it should be possible to reduce the effect by defocusing the telescope and hence increasing the size of the stellar light beam in the fibre. Another solution would consist in optimising HARPS guiding system with, for example, a tip-tilt system \citep{Bouchy99}. 

\subsection{Echelle diagram}

In solar-like stars, p-mode oscillations are expected to produce a characteristic comb-like structure in the power spectrum with mode frequencies $\nu_{n,l}$ approximated reasonably well by the simplified asymptotic relation \citep{Tassoul80,JCD88} :
\begin{equation}\label{asympt}
\nu_{n,l}  \approx (n+\frac{l}{2}+\epsilon)\Delta\nu_0 -  l(l+1) D_0\mathrm{.}
\end{equation}

The two quantum numbers $n$ and $l$ correspond to the radial order and the angular degree of the modes. If the stellar disk is not resolved, only the modes with degree $l \lesssim 3$ can be observed. For given order and degree $n$ and $l$, large and small separations are defined respectively by 
\begin{equation}\label{defls}
\Delta \nu_{n,l}=\nu_{n+1,l}-\nu_{n,l}
\end{equation}
and
\begin{equation}\label{defss}
\delta \nu_{n,l}=\nu_{n,l}-\nu_{n-1,l+2}\mathrm{,}
\end{equation}
 where $\Delta\nu_0$ is the mean large separation, i.e. the mean frequency separation between two modes of consecutive radial orders, and is also defined by 
\begin{equation}
\Delta \nu_0 \approx \left[ 2 \int^R_0 \frac{dr}{c}\right]
\end{equation}
with $R$ the stellar radius and $c$ the sound speed. $D_0$ is given by 
\begin{equation}
D_0 = \frac{<\delta\nu_{n,l}>_n}{4l+6}\mathrm{.}
\end{equation}
Frequencies separations, as well as frequencies of individual modes, were previously reported by \citet{Bouchy02} and \citet{Bedding04}. 

\begin{figure}
\includegraphics[width=5.5cm,height=8.5cm,angle=-90]{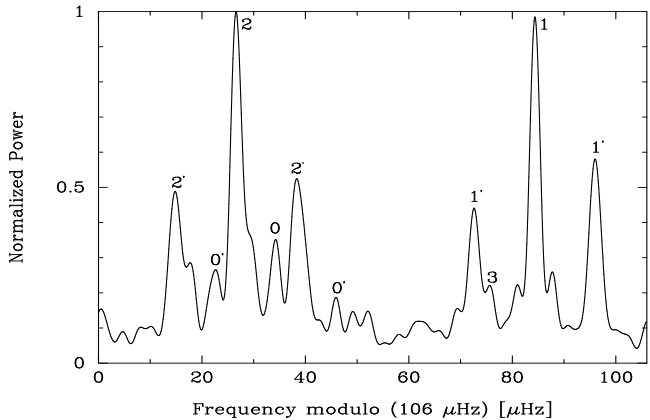}
\caption{Sum of the \'echelle diagram of \acena. Peaks corresponding to the $l=0,1,2\ \mathrm{and}\ 3$ and their daily aliases are identified (with a prime).}
\label{spectrum3}
\end{figure}

The power spectrum shown in Fig.~\ref{spectrum1} presents a clear and unambiguous periodicity of 106 $\mu$Hz both on its autocorrelation or in the comb response, giving an estimate for $\Delta \nu_0$. To identify the angular degree $l$ of each mode individually, we divided the power spectrum into slices of 106 $\mu$Hz and summed them up. The corresponding summed \'echelle diagram is presented in Fig.~\ref{spectrum3} and allows us to unambiguously identify modes $l=0,1,2$ and their side-lobes due to the daily aliases at $\pm$11.57 $\mu$Hz. Some low-amplitude $l=3$ modes were also identified.

\subsection{Identified frequencies}

 We distinguish two different stages in the mode identification process. First we need to extract the frequency from the acoustic spectrum. Roughly speaking this phase consists in locating the high-amplitude peaks in the spectrum. Second we have to verify whether it indeed corresponds to a pulsation mode of the star. We used two different methods to extract the frequencies from the amplitude spectrum, as described below. As a general rule, among the peaks extracted from the spectrum, we selected only the strongest, i.e. with $S/N$ ratios higher than 3 ($\ge$ 11 {\cms}). However, we needed some additional selection process to determine whether each frequency was in acceptable agreement with the asymptotic relation. This kind of process is characteristic of frequency identification in Sun-like stars. It is equivalent to assigning a true or false value to the peaks we can extract from the amplitude spectrum.

 It is worth noting that, regardless of the extraction method used, the spectrum can be treated to improve the mode detection. The coupling between the gap-filling technique and the on-sight extraction, as described below, is an example of such an additional process.

 We first used a CLEAN algorithm \citep{Roberts87}. This method presents the advantage of removing the artifacts in the spectrum resulting from the gaps in the time series. However, after using CLEAN, we still needed to reject some frequencies that were obviously false p-modes identifications. This raised the question of the limits of the CLEAN procedure. The main problem concerns the finite lifetimes of the modes: using CLEAN, we subtract pure infinite sine waves from the signal, despite the fact that the observed modes have a finite lifetime. There is a risk of removing signal that was not originally in the spectrum. 

 The second method supposes an on-sight extraction of the frequencies in the amplitude spectrum. We select each frequency with respect to the template obtained from the summed \'echelle diagram (Fig.~\ref{spectrum3}). Unlike the CLEAN algorithm, it does not modify the original spectrum. We also note that the extraction and frequency selection are simultaneous. The limits of this method are obvious since many structures in the spectrum cannot be distinguished on sight and it cannot account for interferences that may occur between different modes. We also proceeded to an on-sight identification using a pre-treated spectrum with the gap-filling method \citep{Fossat99,Bouchy02}. This method aims at reducing the effect of single-site observation. It consists in replacing gaps in the time series by signals obtained at a time $t=2/\Delta \nu_0$, before or after. It is justified since a peak appears in the autocorrelation spectra for $\Delta \nu_0/2$ \citep{Bouchy02}.

\begin{table}
\begin{minipage}[h!]{\columnwidth}
\caption{P-mode frequencies for {\acena} ($\mu$Hz).}             
\label{freqs}      
\centering           
\setfnsymbol{chicago}              
\renewcommand{\thempfootnote}{\fnsymbol{mpfootnote}}
\renewcommand{\footnoterule}{}
\begin{tabular}{ccccc}        
\hline\hline                 
 & $l=0$ &$l=1$  &$l=2$ &$l=3$ \\    
\hline                        
 $n=16$ &1837.6        &1886.5\footnote{Only with gap-filling}        &1936.6\footnote{Only on-sight (with or without gap filling) \label{gpf}}         &      \\
 $n=17$ &1943.0        &1990.6        &2038.9         &      \\
 $n=18$ &              &2096.8        &2143.9\footnote{Only with CLEAN \label{clean}}         &      \\ 
 $n=19$ &2152.6        &2202.0        &2250.8/2254.4  &2299.4\\ 
 $n=20$ &              &2308.7        &2355.8/2357.6  &      \\ 
 $n=21$ &2363.3        &2415.2\footref{gpf}        &2462.6/2465.0  &2511.5\\ 
 $n=22$ &2470.4        &2519.6        &2568.2\footref{gpf}/2571.8  &      \\ 
 $n=23$ &2576.3        &2626.4        &2674.4/2676.5\footref{clean}  &      \\ 
 $n=24$ &2682.5\footref{clean}        &              &               &2833.8\\ 
 $n=25$ &2788.2\footref{clean}        &2840.1        &               &      \\
 $n=26$ &2895.6\footref{gpf}        &              &               &      \\ 
\hline                                  
\end{tabular}
\end{minipage}	       
\end{table}

\begin{table}
\caption{Amplitudes of the identified p-modes ({\cms}).}             
\label{amps}      
\centering                         
\begin{tabular}{ccccc}        
\hline\hline                 
 & $l=0$ &$l=1$  &$l=2$ &$l=3$ \\     
\hline                        
 $n=16$ &21          &13            &13         &    \\
 $n=17$ &14          &24            &22         &    \\
 $n=18$ &            &32            &12         &    \\ 
 $n=19$ &22          &15            &52/26      &16  \\ 
 $n=20$ &            &37            &13/30      &    \\ 
 $n=21$ &20          &22            &22/23      &14  \\ 
 $n=22$ &19          &48            &25/20      &    \\ 
 $n=23$ &23          &25            &29/18      &    \\ 
 $n=24$ &16          &              &           &16  \\ 
 $n=25$ &13          &14            &           &    \\ 
 $n=26$ &14          &              &           &    \\ 
\hline                                   
\end{tabular}	       
\end{table}
 
 In all cases, $n$ values are assigned to p modes using the asymptotic relation (\ref{asympt}). We assumed that the parameter $\epsilon$ has a value close to the solar one ($\epsilon_{\odot} \sim 1.5$). The frequency and amplitude of the 34 p-modes we identified are listed in Tables~\ref{freqs} and \ref{amps}. Generally the two extraction methods are in good agreement. The idea behind using them was to check their agreement and to obtain complementary sets of frequencies. Therefore, in the following work we used all the frequencies shown in Table~\ref{freqs} and we did not assign weights depending on the detection technique used. However, the frequencies not appearing for all the techniques are marked in Table~\ref{freqs}. For some $l=2$ modes, we identified multiple frequencies.

 Amplitudes were determined by assuming that none of the p modes are resolved and that it corresponds to the height of the peak in the power spectrum after quadratic subtraction of the mean noise level. Considering that the frequency resolution of our time series is 2.6 $\mu$Hz, we adopted an uncertainty on no-resolved oscillation modes of 1.3 $\mu$Hz. Such an uncertainty can be considered as conservative in the case of infinite time-life modes with high $S/N$ ratio ($\ge$3.5). 

\begin{figure}
\includegraphics[width=8.5cm]{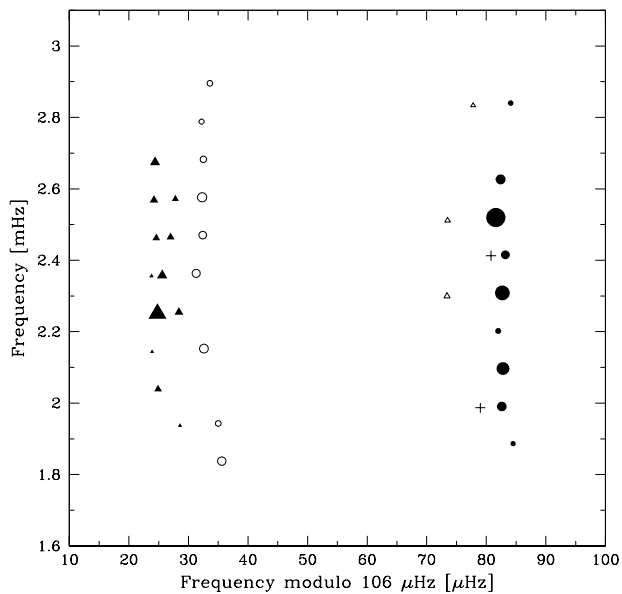}
\caption{Identified p-modes $l$=0 ($\circ$), 
$l$=1 ($\bullet$), $l$=2 ($\blacktriangle$), and $l$=3 ($\triangle$) in \'echelle 
diagram format. The size of the symbols are proportional to the amplitude of the modes. The crosses correspond to ambiguous modes (see text for details).}
\label{echelle1}
\end{figure}

All the identified modes are displayed in the \'echelle diagram in Fig.~\ref{echelle1}. The inferred values for the radial order, $n$, and the angular degree, $l$, are also given in Tables~\ref{freqs} and~\ref{amps}. In some cases, we observed two frequencies corresponding to the same couple $(n,l)$. These multiple frequencies are given for $l=2$. We note that, for each multiple detection at $l=2$, at least one frequency was detected simultaneously with CLEAN and on sight. It therefore indicates that these frequencies are not just the same one identified slightly differently when changing the extraction method. We also extracted peaks at frequencies 1987.0 and 2412.8 $\mu$Hz. The first one cannot be identified precisely as an $l=1$ or $l=3$ mode. The second one is a peculiar case. If we select it, we obtain a multiple detection for an $l=1$ mode. However, the situation is quite different from our $l=2$ cases, since both components here have been identified using a different method. It is thus possible that these two values reflect a difference in the extraction methods used. 
 Therefore, we decided not to include these two frequencies in Table~\ref{freqs}, but we still display them in Fig.~\ref{echelle1}. 

\subsection{Frequency scatter analysis}\label{scatmes}

 If the frequencies were to follow the simplified asymptotic relation given by Eq.~(\ref{asympt}) exactly, their representation in the \'echelle diagram would consist of parallel vertical lines, one for each degree. Although the observations reproduce this general trend, we observe a more complex structure in Fig.~\ref{echelle1}.

 First the asymptotic relation in the form given by Eq.~(\ref{asympt}) is valid for high frequencies, i.e. large compared to the Brunt-V\"ais\"al\"a frequency \citep{Vandakurov67,Tassoul80}. There is a small departure from this relation in the \'echelle diagram. However, for the considered radial orders in the range 16-26, this approximation is accurate. This can be seen in Fig.~\ref{lsep} where the large separation is nearly constant with increasing frequency. 

Second, other physical processes can add some scatter around each individual frequency, $\nu_{n,l}$. In the first instance, because the oscillation modes in sun-like stars are stochastically excited and have finite lifetimes, they appear in the power spectrum as a series of peaks modulated by a Lorentzian. The half width of the Lorentzian reflects the damping rate of the mode and thus its lifetime, see e.g. \citet{Toutain92}. This effect has been well-studied for the Sun, leading to characteristic lifetimes between 38.4 and 3.15 days in the frequency range 1570-2808 $\mu$Hz \citep{Chaplin97}. \citet{Kjeldsen05} suggest that the scatter they measured around the frequencies in {\acena} correspond to this finite lifetime effect. They give the values of 2.3$^{+1.0}_{-0.6}$ days at 2.1 mHz and 2.1$^{+0.9}_{-0.5}$ days at 2.6 mHz.

Furthermore, since the star is rotating, the proper frequency of a given mode will be split, in the observer frame, following the relation 
\begin{equation}
\nu_{n,l,m}\simeq \nu_{n,l} + m\Omega
\end{equation}
where $m$ is the azimuthal order ($-l\leqslant m \leqslant l$) and $\Omega$ the rotation rate. We suppose a rigid body rotation here. In the case of the Sun, \citet{Chaplin01} find an average value of 0.4 $\mu$Hz for the rotational splitting at low degrees ($l \leqslant 3$). Following their detection of p-modes in \acena, \citet{Bouchy02} suggest that they could have observed hints of a rotational splitting of the order on 0.5 $\mu$Hz, compatible with an almost solar rotational period.

\begin{figure}
\includegraphics[width=8.5cm]{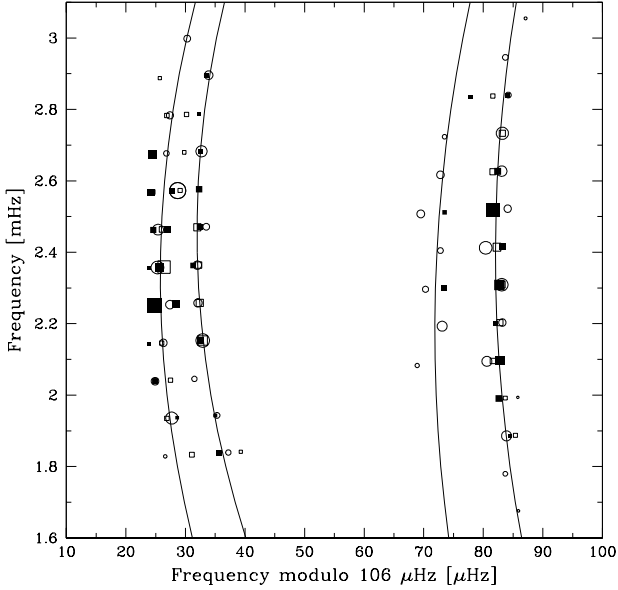}
\caption{Echelle diagram for all the identified frequencies in {\acena} provided with HARPS (black-square), CORALIE (open-square) and UVES-UCLES (open-circle). The lines correspond for each angular degree to the fit of a second-order polynomial to all frequencies. We used them as guidelines to compare the different sets of frequency.}
\label{echelle2}
\end{figure}

 Scatter analysis is a very interesting tool, especially when dealing with short seismic runs that do not provide a frequency resolution high enough to allow more direct techniques, such as direct fitting of the acoustic power spectrum \citep{Gizon03,Ballot06,Fletcher06}. To measure the scatter, we first needed to estimate the departure from the asymptotic relation for each frequency, since we used it as a reference for the mode identification. For this purpose, we fitted the frequencies, for a given degree, with second-order polynomials. This method is close to the one used by \citet{Kjeldsen05}. However, since we observe multiple frequencies (and do not average them) for some given $(n,l)$ values, we could not use the same variables as they did (see their Eqs.~(2)-(5)). We instead adopted the variables displayed in the echelle diagram, i.e. the frequency and the frequency modulo the average large separation, the latter one being a second-order polynomial function of the former one.

 The scatter $\sigma^{(l)}$ on the frequencies for a given degree $l$, with respect to the corresponding polynomial fit, is defined as the square root of the normalised weighted sum of the fit residuals:
\begin{displaymath}\label{rmsscat}
\sigma^{(l)}=\left( \frac{N_l\displaystyle\sum_{i=1}^{N_l} a_i(f^{mod}_{i,obs}(l)-f^{mod}_{i,fit}(l))^2}{(N_l-3)\displaystyle\sum_{i=1}^{N_l} a_i} \right)^{\frac{1}{2}}
\end{displaymath}
with $f^{mod}_{i,obs}(l)$ the $i$th frequency observed at degree $l$ modulo $\Delta\nu_0$, $a_i$ the corresponding amplitude, $f^{mod}_{i,fit}(l)$ the fitted value, and $N_l$ the total number of frequencies observed at $l$. As a first approximation, we consider that the amplitudes of the modes represent suitable weights.

Assuming normally distributed frequencies around the fitting polynomial, the uncertainties on the computed scatter can be approximated by $\sigma^{(l)} / \sqrt{2N}$, with 
$N$ the number of points used for the computation. 
 The resulting scatters for individual degrees are\footnote{In the following, the subscript HRP will stand for HARPS, COR for CORALIE, and UVC for UVES/UCLES} $\sigma^{(0)}_{\mathrm{HRP}} = 0.50 \pm 0.12$~{\muhz}, $\sigma^{(1)}_{\mathrm{HRP}} = 0.70 \pm 0.17$~{\muhz}, and $\sigma^{(2)}_{\mathrm{HRP}} = 1.71 \pm 0.34$~{\muhz}.

 The $l=3$ modes were excluded from our analysis. We only observed three of them and, thus, cannot derive a proper scatter using a second-order polynomial fit. All these modes have low amplitudes, so discarding them should not alter our analysis.

 As discussed in Sect.~\ref{spltchar}, this scatter increase with mode degree can be a consequence of rotational splitting. We searched for such trends in the previously published sets of frequencies. In Fig.~\ref{echelle2} we represent the values from the previous papers in the \'echelle diagram, in addition to our frequencies. The lines appearing in this \'echelle diagram are fits to all the frequencies and we used them as visual guidelines for the comparison between frequency sets. It is worth stressing that the general agreement between these three sets is very good.

\paragraph{CORALIE frequencies.}

  From the published frequencies, the scatters are $\sigma^{(0)}_{\mathrm{COR}} = 0.96 \pm 0.24$~{\muhz}, $\sigma^{(1)}_{\mathrm{COR}} = 0.90 \pm 0.20$~{\muhz}, and $\sigma^{(2)}_{\mathrm{COR}} = 1.01 \pm 0.23$~{\muhz}. These show a marginal increase with degree $l$. However, comparing the three sets of frequencies in Fig.~\ref{echelle2} suggests that the identifications in \citet{Bouchy02} for  the $l=0$ modes at frequencies 2573.1 $\mu$Hz, 2679.8 $\mu$Hz, and 2786.2 $\mu$Hz should be modified. We assign them a degree $l=2$. The resulting scatters are $\sigma^{(0)}_{\mathrm{COR}} = 0.33 \pm 0.11$~{\muhz}, $\sigma^{(1)}_{\mathrm{COR}} = 0.90 \pm 0.20$~{\muhz}, and $\sigma^{(2)}_{\mathrm{COR}} = 1.53 \pm 0.30$~{\muhz}. In this case the trend towards increasing at higher degrees clearly appears. This is not surprising since we obtain two possible new multiple frequencies for $l=2$, at radial orders 22 and 24, when changing the identification.

\paragraph{UVES/UCLES frequencies.} The picture for the UVES/UCLES frequency set is less clear. It should first be noted that multiple frequencies were also found for certain $(n.l)$ values in \citet{Bedding04}. Nevertheless, in order to retain only one frequency, those multiple values were averaged using a weighted-mean. We also note that, based on their ridge identification technique, they retained modes with $S/N < 3$. In order to compare homogeneous sets of frequencies, we only retain their modes with $S/N > 3$ (i.e. with amplitudes larger than 8.7 {\cms}). As a first approximation we can consider that the effect of averaging the multiple frequencies is negligible. To support this assumption, we raise the fact that all the multiple frequencies identified are composed of a high-amplitude and a low-amplitude peak. The weighted mean is dominated by the latter one. Furthermore, the mode at 2572.7~{\muhz} has been identified as both $l=0$ and $l=2$ in their Table~1. Comparing with the HARPS and CORALIE frequencies in Fig.~\ref{echelle2}, we choose to retain only the second identification. The resulting scatters are $\sigma^{(0)}_{\mathrm{UVC}} = 1.04 \pm 0.25$~{\muhz}, $\sigma^{(1)}_{\mathrm{UVC}} = 1.32 \pm 0.28$~{\muhz}, and $\sigma^{(2)}_{\mathrm{UVC}} = 1.32 \pm 0.28$~{\muhz}. The trend is much more marginal in this case and only visible between $l=0$ and $l=1$ scatters. The fact that multiple frequencies are observed for both $l=1$ and $l=2$ modes may be the main reason for this lack of increase. This may be coupled to resolution effects (see discussion in Sect.~\ref{spltchar}). The same kind of behaviour is observed if we compute the scatters from the non-averaged frequencies (T. Bedding, private communication), $\sigma^{(0)}_{\mathrm{UVC}} = 1.13 \pm 0.27$~{\muhz}, $\sigma^{(1)}_{\mathrm{UVC}} = 1.84 \pm 0.38$~{\muhz}, and $\sigma^{(2)}_{\mathrm{UVC}} = 1.77 \pm 0.36$~{\muhz}. We should eventually point out that if we compute the scatter while retaining all the published frequencies (i.e. without cutting at $S/N > 3$), it decreases between $l=1$ and $l=2$ modes.

 The above discussion shows that, for the HARPS and CORALIE frequency sets, the scatter shows a clear trend towards increasing with the mode degree. It is not possible to distinguish such a trend in the UVES/UCLES frequencies, although the large uncertainties on the scatters for $l=1$ and $l=2$ do not exclude it. It is also worth stressing the sensitivity of the scatter determination to the mode identification, as shown with the CORALIE and UVES/UCLES frequencies. This observed trend could nevertheless help for characterising the multiple frequencies observed for $l=2$ modes. 

\subsection{Characterisation of the split modes}\label{spltchar}

 The $m$-dependent visibility of a rotationally split mode $(l,m)$, independently of the radial order $n$, is linked to the inclination $i$ of the rotation axis of the star by the coefficient \citep{Gizon03} :
\begin{equation}\label{epsrot}
\mathscr{E}_{lm}(i)=\frac{(l-|m|)!}{(l+|m|)!}\left[ P^{|m|}_l (\cos i) \right]^2 
\end{equation}
where $P^{|m|}_l$ is the Legendre polynomial. 

  According to Eq.~(\ref{epsrot}) at high inclinations (i.e. $i \gtrsim 65 \degr$), modes with the highest visibility coefficients $\mathscr{E}_{lm}(i)$, and thus the highest amplitudes, are those with $l=1$, $m=1$ and $l=2$, $m=2$. Modes with $l=2$, $m=0$ could possibly be seen, but their amplitude in the power spectrum is expected to be approximately two times lower than their $m=2$ counterparts. Modes with $l=1$, $m=0$ and $l=2$, $m=1$ almost disappear for high inclinations.

 In the framework of the rotational splitting hypothesis, we can interpret the multiplets found in Table~\ref{freqs} for some $l=2$ modes as rotationally induced. This assumption is justified by the above discussion. If we make the assumption that the inclination axis of {\acena} is aligned with the axis of the binary system, we can then consider $i=79\degr$ \citep{Pourbaix99}. It follows that modes with $(l,m) = (1,1)$ and $(2,2)$ should be the most visible. Whereas the effect of mode lifetimes on the scatter should not depend on the mode degree, an enhancement with increasing $l$ and, therefore, $|m|$ is possible, explaining the trend observed in HARPS and CORALIE scatters. We eventually note that assigning azimuthal orders to the split modes we observe does not exclude a contribution coming from the finite mode lifetime effect to the overall scatter, as discussed in the following.

\begin{figure}
\includegraphics[width=5.5cm,height=8.5cm,angle=-90]{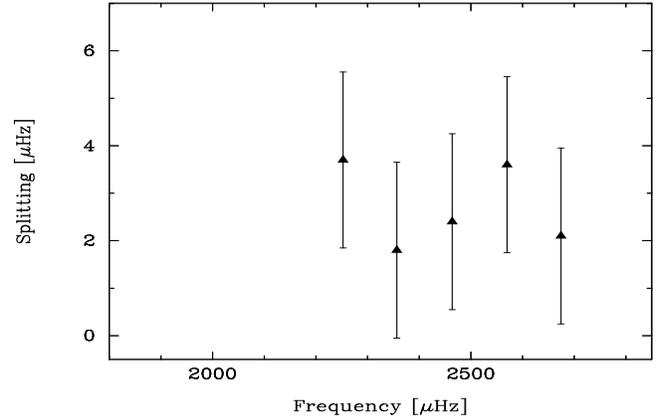}
\caption{Measured splitting for $l=2$ modes in HARPS frequencies. Symbols are the same as in Fig.~\ref{echelle1}.}
\label{rsplit}
\end{figure}

  The measured separations in our frequency multiplets at $l=2$ are plotted in Fig.~\ref{rsplit}. The error bars have been calculated using the quadratic sum of the uncertainties on our frequencies. The largest possible splitting effect would be obtained for modes with $m=\pm 2$. Assuming an almost solar rotation rate, the difference would be around 2 $\mu$Hz, which is compatible with the values displayed in Fig.~\ref{rsplit}.
 However, the diagnostic potential of these separations should be considered cautiously. On one hand, if the mode lifetimes are long compared to the observing time, $T$, then the accuracy to which two close frequencies can be resolved is limited by our resolution, which scales approximately like $1/T$. Depending on the phase difference between the modes, they could result in only one peak in the power spectra or, on the contrary, two peaks separated from $\gtrsim 1/T$, even if the true distance between them is smaller than the measured splitting \citep{Loumos78,JCD82}. This could lead to overestimate the rotational splitting. We note that, under the above hypothesis on the mode lifetimes, this resolution effect could be involved in the large scatter we observed in the UVES/UCLES frequencies at $l=1$. This effect certainly implies that a small distance in frequency in the power spectrum, caused by splitting for $l=1$ modes, will be more severely biased than a larger one, caused in this case by an $l=2$ splitting. On the other hand, if the mode lifetimes are similar to or shorter than the observing time, their contribution to the scatter could become similar to or greater than the effect of rotation. In this case, it becomes difficult to disentangle the two contributions and heavier simulations are needed.

 \citet{Gizon03} and \citet{Ballot06} used simulated seismic data, spectra corresponding to long time series of respectively 180 and 150 days, to study the possibility of estimating the rotation rate and the inclination of a sun-like star. The spectra models used to describe these simulations were assigned $\Omega$ and $i$ as the only two free parameters. These were estimated using a maximum-likelihood method. \citet{Gizon03} consider a single-split mode in the spectrum, whereas \citet{Ballot06} use a multi-mode fitting approach. In relation to this study, the main conclusion from both studies is that, for a rotation rate $\Omega = \Omega_{\odot}$, it is not possible to determine accurately the inclination that introduces a bias in the estimated rotational splitting. Moreover, because of computational costs, their studies were made with fixed mode lifetimes. It is not clear which degree of complexity would be added to this problem if mode-lifetime estimation were to be included. Given the length of the time series considered in these works, it is doubtful that a conclusion could be reached using short seismic runs such as the one presented here. Still, this gives a good insight into the overall problem of estimating stellar parameters from acoustic spectra of solar-like stars. It also points out the importance of having an independent estimate of the inclination, as in our case, in order to estimate a rotation rate using asteroseismology.

 \citet{Kjeldsen05} estimated mode lifetimes in {\acena} using the UVES/UCLES frequencies. Their method consists in simulating several realizations of a time series corresponding to their observations and from them extracting frequencies to obtain a theoretical estimate of the scatter. They needed to compute at least 100 simulations per set of input parameters. However, they supposed a static star in their study, and thus neglected rotation. Using this technique with an additional parameter to estimate is possible, although it would represent several months of CPU time. Describing possible methods to bypass this prohibitive computing time is beyond the scope of this paper and will be addressed in a later study. Another consequence of our identification of splitting is that neglecting rotation may lead to underestimating mode lifetimes, since they should not account for all the observed scatter. The value given for the mode lifetimes in \citet{Kjeldsen05} should thus be considered as a lower limit.

 Independent measurements of the rotation of {\acena} have been made, which could indicate the expected splitting.
 A rotational period of $28.8\pm 2.5$ days has been measured for {\acena} from analysis of the modulation of {\it IUE} spectra \citep{Hallam91}. However, this paper states that several solutions come out of the analysis. Although the result they give is their best estimate, they are not able to definitely rule out other shorter rotational periods. The splitting corresponding to this rotational period would be $0.40 \pm 0.04$~{\muhz}. \citet{Saar97} derived a projected velocity, $v\sin i =2.7 \pm 0.7$ km.s$^{-1}$, from high-resolution observations of FeI, CaI and NiI lines. Using the value of the radius determined by interferometry, $R=1.224\pm 0.003\ R_{\odot}$ \citep{Kervella03} and the inclination from \citet{Pourbaix99}, we obtained $P = 22.5 \pm 5.9$ days from this measurement and a corresponding splitting $\Omega = 0.51 \pm 0.13$~{\muhz}.

 More recently, \citet{Fletcher06} used a 50-day time series from the WIRE satellite to estimate the rotational splitting and mode lifetimes of {\acena}. They used both power spectrum  and autocovariance fitting techniques. Their best estimate gives a rotational splitting of $0.54 \pm 0.22$~{\muhz} and a mode lifetime of $3.9 \pm 1.4$ days. This value is higher than the one derived by \citet{Kjeldsen05}, which supports our analysis stating that the mode lifetime should not be the sole contributor to the frequency scatter and that neglecting the splitting should introduce a significant bias.

\subsection{Large and small separations}

\begin{figure}
\includegraphics[width=5.5cm,height=8.5cm,angle=-90]{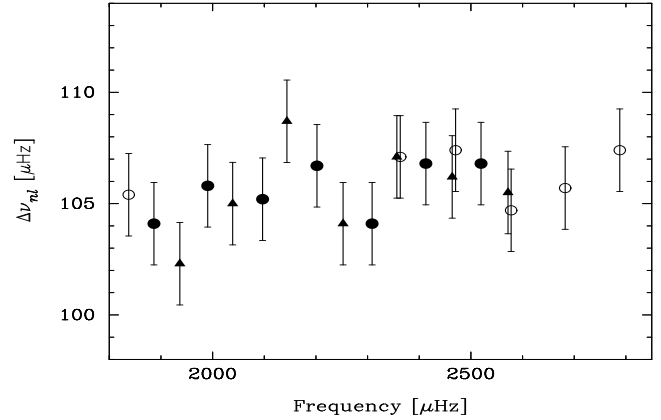}
\caption{Large separations $\Delta\nu$ obtained from HARPS frequencies versus frequency for p-modes of degree $l$=0, 1 and 2. Symbols are the same as in Fig.~\ref{echelle1}.}
\label{lsep}
\end{figure}

\begin{figure}
\includegraphics[width=5.5cm,height=8.5cm,angle=-90]{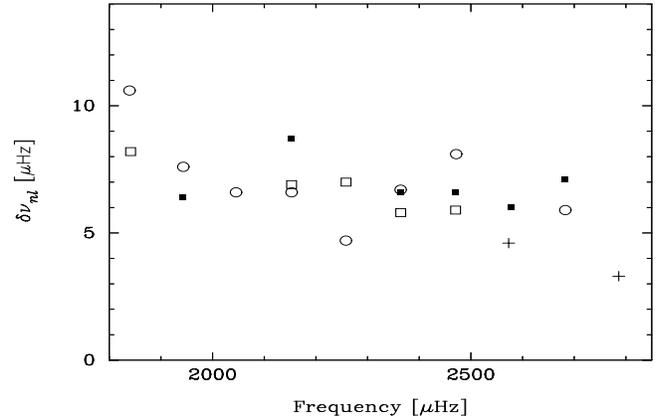}
\caption{Small separations $\delta\nu_{nl}$ for $l=0$. The different symbols correspond to the values obtained with HARPS (black-square), CORALIE (open-square) and UVES-UCLES (open-circle). Crosses mark the two CORALIE small separations which disappear after the new identification suggested in Sect.~\ref{scatmes}. For sake of clarity the error bars were omitted.}
\label{ssep}
\end{figure}

Individual large and small separations obtained from the new HARPS frequencies are represented in Fig.~\ref{lsep} and \ref{ssep} and uncertainties on both of them are 1.8~{\muhz}. For $l=2$ modes with multiple frequencies detected, we used the mean frequency value in Eqs.~(\ref{defls}) and (\ref{defss}). The small separations are very useful for stellar modelling and need to be strongly constrained, e.g. \citet{Montalban06}. The first two HARPS small-separation values were derived using $l=2$ modes with only one frequency. It should be emphasised that, considering rotational splitting, these values are questionable since one $m=\pm 2$ mode could be missing.

 Fig.~\ref{ssep} also displays the small separations obtained from the CORALIE and UVES/UCLES runs. We note that the HARPS values do not show the trend of CORALIE small separations towards decreasing sharply at high frequencies. In that respect, they show very good agreement with the separations derived from the UVES/UCLES run. \citet{Bedding04} also identify some modes with multiple frequencies. They compute the corresponding separations using a weighted mean (using the amplitudes as weights). Although their treatment differs slightly from ours, we consider it a small effect that does not challenge the overall agreement between HARPS and UVES/UCLES small separations. Eventually, the new identification suggested in Sect.~\ref{scatmes} for the CORALIE frequencies results in the disappearance of the last two small separations at 2573.1~{\muhz} and 2786.2~{\muhz} for this set of frequencies. We note that this improves the agreement between the three runs. 

 We note that averaging without weighting our $l=2$ frequencies may lead to underestimate the contribution to the frequency scatter coming from the mode lifetime. On the other hand a weighted average is unnecessary for rotationally-split modes since their contribution to the mean frequency should be the same when averaging over $m$. The fact that the HARPS and UVES/UCLES small separations agree well shows that the assumption made before averaging does not have a huge effect on the computed values. The important point is therefore to obtain more than one frequency, at least for $l=2$, so that both effects can be taken into account by using mean frequency values.

  Furthermore, from the results displayed in Figs.~\ref{lsep} and~\ref{ssep}, we obtained new mean values for the large and small separations for the HARPS frequencies, respectively, $105.9 \pm 0.3$ $\mu$Hz and $6.9 \pm 0.4$ $\mu$Hz. The latter value was obtained by averaging the small separations over $n$. The former was estimated by linearly fitting the relation $\nu_{n,l}=n\Delta \nu_0 + k$, because consecutive values of the large separation are not independent; here $k$ is a constant and the slope is the average large separation. Using the same methods, we obtained $105.5 \pm 0.3$~{\muhz} and $6.8 \pm 0.4$~{\muhz} for CORALIE (using the new identifications given in this paper) and $106.1 \pm 0.4$~{\muhz} and $7.1 \pm 0.6$~{\muhz} for UVES/UCLES. These values stress the very good agreement existing between the different runs.

\section{Conclusion}

 We analysed observations of {\acena} made during a five-night run using the high-precision spectrograph HARPS. We extracted 34 frequencies from the amplitude spectrum, confirming the previous studies of this star and were able to improve the characterisation of the observed p-modes.

 Our main result came from analysing the scatter in these frequencies. We found clear evidence of rotational splitting in the spectrum of {\acena}: an increase in the scatter with the angular degree of the modes and multiplets for $l=2$ modes. This observation suggests that the value of the mode lifetime should be higher than the one given by \citet{Kjeldsen05} without taking the rotation into account. We thus conclude that a frequency scatter analysis must include the rotational splitting effect.

 Observing rotational splitting allows us to derive precise values for the large and small separations, using frequencies averaged over the azimuthal order $m$. These new values are in good agreement with the small separations obtained from the UVES/UCLES run \citep{Bedding04} and contradict the trend seen in those obtained from CORALIE \citep{Bouchy02}. This result is fundamental for further modelling of {\acena}. 

 Although we argue against short mode lifetimes, this problem still requires more observing time, on longer periods, to be fully addressed. Such observations would also allow us to obtain a better resolution, which is necessary for detecting very close frequencies unambiguously and obtaining direct measurements of the rotational splitting.  

 This study shows that ground-based, single-site, seismic observations are valuable for improving our knowledge of the pulsation modes in sun-like stars.
 Stronger seismic constraints are also needed on {\acenb} in order to model the entire system efficiently. These results are also very encouraging in the framework of future base-ground experiments. It is highly relevant that rotational splitting could be measured even from single-site measurements and demonstrates the strong possibilities offered by asteroseismology, especially if longer times become available.

\begin{acknowledgements}
F. Pepe and G. Locurto are acknowledged for their fruitful discussion and help about the guiding noise. We thank J. Christensen-Dalsgaard and our referee for their precious advice and comments.

\end{acknowledgements}

\bibliography{5694ref}

\end{document}